\documentstyle[12pt]{article}
\title{POST-NEWTONIAN FRAMES OF REFERENCE}
\author{Ll. Bel \\
Lluis.Bel@Wanadoo.fr}
\date{\today}
\begin{document}
\maketitle
 
\begin{abstract}
 
A general theory of frames of reference proposed in a preceding
publication is considered here in the framework of the post-Newtonian
approximation, assuming that the frame of reference is centered on a
time-like geodesic. The problem of taking into account the rotation
of the frame of reference, which is usually ignored or incorrectly
oversimplified, is here discussed in detail and solved. 
 
\end{abstract}
 
\section{Introduction}
 
The problem of defining Post-Newtonian reference frames and
developing the corresponding theory of adapted coordinate
transformations has been discussed in two main papers: \cite
{Brumberg-Kopejkin}, and \cite {Damour-Soffel-Xu} in the framework of
covariant metric theories of gravity and related fields, and by \cite
{Klioner-Soffel} in one of the so-called PPN frameworks. None of
these papers is based on a general theory and they all suffer of a
long-lasting confusion between frames of reference and systems of
coordinates. As a consequence they rely more on
improvisations suggested on the way by details of the approximate
calculations than on a well-defined logical line.
The first paper considers frames of reference with
rotation but it does it in an oversimplified and insufficiently accurate
way.
The two other papers ignore the rotating frames of reference and
are for this reason of restricted interest as far as the theory of
such frames is concerned..
 
In \cite {Aguirre-al} and \cite {Bel-94} a general theory
of frames reference and a particular post-Newtonian application was
considered. This tentative theory has been subsequently modified and
a few other applications have been reviewed in \cite {Bel-99}. This paper
deals with an explicit description of this renewed theory in the
framework of the post-Newtonian approximation.
 
The second section contains a short account of the main ingredients
built in the concept of a frame of reference, namely the definition
of a meta-rigid motion and a chorodesic synchronization. It includes
also the definition of a particular type of frames of reference: the
quo-inertial ones. The third section describes the general framework
of the post-Newtonian approximation, slightly enlarged to include all
types of inertial fields at the Newtonian approximation. The fourth
and fifth  sections deal with the main object of this paper: to use
the general definition of Sect. 2 to develop a restricted but
explicit theory of frames of reference as as far as the
post-Newtonian approximation allows it. The Appendix contains a
summary of a few basic definitions.
 
\section{Frames of Reference}

A frame of reference is a pair of geometrical objects
intrinsically defined in the space-time being considered:
\begin{itemize}
 
\item A time-like congruence $\cal R$ of a particular type, that we
shall call the {\it Meta-rigid motion} of the frame of reference,
 
\item A space-like foliation of a particular type $\cal S$, that we
shall call the {\it Chorodesic synchronization} of the frame of
reference.,
\end{itemize}
 
\subsection{\it Meta-rigid motions}.

A meta-rigid motion $\cal R$
has to possess at least three general properties
which are meant to make this concept to inherit as much as
possible of the formal properties of rigid motions in classical
physics:
 
i) Their local characterization must be intrinsic and have a
meaning independently of the space-time being considered.
 
ii) The knowledge of any open sub-bundle of the congruence
must be sufficient to characterize the whole congruence.
 
iii) Its definition does not
have to distinguish any particular world-line of the congruence.
 
To implement these three properties the most natural
approach is to require the vector field $u^\alpha$ of the
meta-rigid motions to satisfy a set of differential conditions.
Obvious candidates which satisfy this requirement are the differential
equations defining the Killing congruences\,\footnote{Definitions in
the Appendix.}:
 
\begin{equation}
\label {0.10}
\Sigma_{\alpha\beta}=0, \quad
\partial_\alpha\Lambda_\beta-\partial_\beta\Lambda_\alpha =0.
\end{equation}
 
Other congruences to be acceptable as meta-rigid motions
are the Born congruences, which are a generalization of
the Killing congruences and are defined by the single group of
conditions:
 
\begin{equation}
\label {0.11}
\Sigma_{\alpha\beta}=0.
\end{equation}
 
We shall continue to call these congruences rigid congruences, as usual.
Meta-rigid ones will be an appropriate generalization of them.
 
The Born conditions, and a fortiori the Killing conditions, are very
restrictive in any space-time including Minkowski's one
(\cite{Herglotz}, \cite{Noether}), and the question may be raised of
generalizing the concept of meta-rigid motion in such a way that the
family of congruences $\cal R$ generalize Born congruences.
 
When facing this problem confronted with a physical application which
requires the sublimated modelisation of the behavior of a rigid body,
many authors resort to use Fermi congruences. By this we mean the
time-like congruences which accept adapted Fermi coordinates based on
a distinguished world-line seed. Because of this distinction Fermi
congruences are not in general acceptable candidates to the concept
of a meta-rigid motion.   
 
Harmonic congruences can be defined as those time-like congruences,
with unit tangent vector $u^\alpha$, for which there exist three
space-like functions $f^a(x^\alpha)$ satisfying the following
equations:
 
\begin{equation}
\label {0.12}
\Box f^a=0, \quad
u^\alpha\partial_\alpha f^a=0, \quad a,b,c=1,2,3
\end{equation}
where $\Box$ is the d'Alembertian operator corresponding to
the space-time metric. Harmonic congruences, with the corresponding
harmonic coordinates,
have been used extensively in the literature for several reasons,
including mainly the fact that they simplify many calculations.
Harmonic congruences could in principle be considered good
representatives of meta-rigid motions.  In fact they are not. They
are acceptable generalizations of Killing congruences but they are
not generalizations of the Born congruences. More precisely it has
been proved \cite{Bel-Coll} that irrotational, and not Killing, Born
congruences are never harmonic congruences.
 
To solve the problem just mentioned concerning the harmonic
congruences it has been proposed to identify the meta-rigid motions
with appropriately selected {\it Quo-harmonic congruences}
(\cite{Salas}, \cite{Bel-Llosa1}, \cite{Bel-Llosa2}). These being
those congruences for which there exist three independent space-like
solutions $f^a(x^\alpha)$ satisfying the following equations:
 
\begin{equation}
\label {0.15}
\hat\triangle f^a \equiv 
(\Box + \Lambda^\alpha\partial_\alpha)f^a=0, \quad
u^\alpha\partial_\alpha f^a=0, \quad a,b,c=1,2,3
\end{equation}
or equivalently, using adapted coordinates and the notations of the
Appendix:
 
\begin{equation}
\label {0.15bis}
\hat\triangle f^a=\hat g^{ij}(\hat\partial_i\hat\partial_j
-\hat\Gamma^k_{ij}\hat\partial_k)f^a=0, \quad f^a=f^a(x^i)
\end{equation}
 
These are proper generalizations of Born congruences in the sense
that any Born congruence is also a quo-harmonic one. In fact when
using adapted coordinates the operator $\hat \triangle$ in the Eq.
above becomes the usual Laplacian operator corresponding to a
3-dimensional Riemannian metric.
 
Besides the above mentioned general local conditions to be imposed to
the concept of {\it Meta-rigid motions} global specific conditions
will be necessary depending on the particular global structure of the
space-time being considered. Also particular but important will
be the case, where the congruence contains a time-like geodesic and
it make sense, physically, to implement the principle of geodesic
equivalence.
 
Notice that quo-harmonic congruences can be described and studied
using any system of coordinates.  But {\it quo-harmonic
coordinates}, i.e. any system of three independent functions $f^a$
of Eq. \ref {0.15}, are in general the most convenient ones to use. 
 
\subsection{\it Chorodesic synchronizations}.
 
The choice of a foliation to define the synchronization time of a
frame of reference is to some extent less essential than the choice
of a quo-harmonic congruence to define its meta-rigid motion.
Besides, time is a much better understood concept in General
relativity than the concept of space. The role of a synchronization
is to be a first step towards a convenient definition of a universal
scale of time, and the properties that are to be required from a
synchronization will depend on the use for which it is intended.
Atomic and sidereal time are two scales of common use which
correspond to different synchronizations of the frame of reference
co-moving with the Earth. Not to mention many other scales of time
used in astronomy. On the contrary, for instance, it would not make
sense to use different types of congruences to describe the motion of
the Earth, at the approximation which considers it as rigid.

We consider here briefly how the {\it International Atomic Time
(IAT)} scale is defined. It is based on the definition of the second
in the international system of units {\it SI} as a duration derived
from the frequency of a particular atomic transition.  
This definition is universal, in the sense that any physicist is able
to use a well defined second, but at the same time is local because
the identification of atomic time with proper-time duration implies
that this definition of the second can be only used to define a scale
of time along the world-line of the clock being used as a
standard reference.
 
To define a scale of time on Earth to be used on navigation systems
like the GPS for instance requires a different approach. In this
case\,\footnote{See for instance \cite{Soffel}} the second is defined
as above with some precisions added. The resulting definition of {\it
IAT} is:
 
{\it IAT is a coordinate time scale defined in a geocentric reference
frame with an SI second realized on the rotating
geoid}.\,\footnote{The geoid is a level surface of constant
geo-potential (gravity and centrifugal potential) at ``mean sea
level", extended below the continents.}
 
It follows from this that the operational definition of
the second is relative to some particular locations, i.e. world-lines
of a particular congruence, which in this case\, \footnote{If the
influence of the Sun and the Moon is neglected} is the rotating Killing
congruence corresponding to the frame of reference co-moving with the
Earth. A second at some other location is then defined as to nullify
the relativistic red-shift between any pair of clocks of reference
co-moving with the Earth. Therefore the second at any other location, not
in
the geoid, is then the interval of time separating the arrival of
two light signals sent, one second apart, from a standard clock
located on the geoid. The red-shift formulas of General relativity
allow then to compare the rhythm of any two clocks that can be joined
with light signals.

Let $\cal R$ be any time-like congruence. A {\it Chorodesic} $C$ of $\cal
R$ is by definition, \cite{Salas} \cite{Bel-Llosa2}, a space-like line
such that its tangent vector $p^\alpha$
satisfy the following equations:
 
\begin{equation}
\label{0.53}
\frac{\nabla p^\alpha
}{d\lambda } =\frac 12u^\alpha \Sigma _{\mu \nu }p^\mu p^\nu, \quad
p^\alpha=\frac{dx^\alpha }{d\lambda }
\end{equation}
where $\lambda$ is the proper length along $C$.
Obviously if $\Sigma _{\mu \nu }=0$, i.e.
if $\cal
R$ is a Born congruence then any chorodesic of $\cal R$ is also
a geodesic of the space-time.
 
Chorodesics of a congruence are important mainly because of the
following result:
 
{\it If a space-like chorodesic $C$ is orthogonal to a world-line of
a congruence $\cal R$ then it is orthogonal to all the world-lines of the
congruence $\cal R$ that it crosses.}
 
This follows from differentiating the scalar product $p^\rho u_\rho$
along
$C$. We thus get:
 
\begin{equation}
\label {0.50}
\frac{d}{d\lambda}(p^\rho u_\rho)=
-\frac 12\Sigma _{\mu \nu }p^\mu p^\nu+
\frac12 p^\rho p^\sigma(\nabla_\rho u_\sigma+\nabla_\sigma u_\rho) 
\end{equation}
and using the definition of $\Sigma _{\mu \nu }$ this is equivalent
to:
 
\begin{equation}
\label {0.51}
\frac{d}{d\lambda}(p^\rho u_\rho)=-2(p^\rho u_\rho)(p^\rho \Lambda_\rho)
\end{equation}
from where the statement above follows.
 
Particular foliations associated with a congruence $\cal R$ are the
one parameter family of hyper-surfaces generated by the
chorodesics orthogonal to any particular world-line of the
congruence. We call these foliations {\it Chorodesic
synchronizations.} In general a chorodesic synchronization depends on
a world-line seed, except if the congruence is integrable in which
case all its
chorodesic synchronizations coincide with the family of hyper-surfaces
orthogonal to the congruence.
 
Let $\cal S$ be the chorodesic synchronization orthogonal to some
world-line $W$ of a frame of reference $\cal R$. Then the associated
{\it Atomic (or Proper) Time Coordination (ATC)} scale is by
definition a time coordinate which value at any event $E$ is the
proper time interval between some arbitrary event on $W$ and the
intersection of $W$ with the leave of $\cal S$ which
contains $E$.
 
Notice that, even when the congruence of reference $\cal R$ is
hypersurface orthogonal, i.e. there exists a single chorodesic
synchronization, the {\it ATC} may depend on the particular
world-line $W$ which is used to identify the coordinate time with
proper-time along $W$.
 
Switching from a world-line seed, $W$, of an {\it ATC} to another
$W^\prime$ gives rise to a sub-group of adapted coordinate
transformations\, \footnote{An analytical example of such sub-groups
can be seen in \cite {Salas}}.  A complete definition of the {\it ATC}
requires therefore to choose a particular world-line $\cal S$, or
eventually a bunch of equivalent ones. On Earth the {\it ATC}, i.e.
the {\it IAT}, is associated to the bunch of world-lines $\cal S$ of
locations on the geoid.    
 
The interest of the consideration of chorodesic synchronizations
stems from the fact that they generalize the hyper-surfaces
orthogonal to integrable congruences, as well as being a particular
case of synchronizations
of equal cyclic adapted time for both static and stationary
space-times.
 
\section{Inertial and gravitational fields: The Post-Newtonian
framework}
 
Our general framework will consist of the following items:
 
\begin{itemize}
 
\item A domain $D$ of $R_4$ to be called the primary domain, or
global domain if global conditions are required.
   
\item A four dimensional space-time metric with signature $+2$,
$g_{\alpha\beta}$, which can be expanded in power of $1/c$ as follows:
 
\begin{eqnarray}
\label {1.1}
g_{00}&=&-1+\frac{1}{c^2}f_{00}+\frac{1}{c^4}h_{00} \\
g_{0j}&=&\frac{1}{c}f_{0j}+\frac{1}{c^3}h_{0j} \\
g_{ij}&=&\delta_{ij}+\frac{1}{c^2}h_{ij}
\end{eqnarray}
The corresponding developments of $\xi, \varphi_i$, and the Fermat
quo-tensor $\hat g_{ij}$ are:
 
\begin{eqnarray}
\label {1.2}
\xi&=&1-\frac{1}{2c^2}f_{00}-\frac{1}{2c^4}(h_{00}+\frac{1}{4}f_{00}^2)
\\
\varphi_i&=&\frac{1}{c}f_{0i}+\frac{1}{c^3}(h_{0i}+f_{0i}f_{00}) \\
\hat g_{ij}&=&\delta_{ij}+\frac{1}{c^2}\alpha_{ij} \quad
\alpha_{ij}=h_{ij}+f_{0i}f_{0j} 
\end{eqnarray}
Every further term will be systematically neglected.
\end{itemize}
 
Our general requirements on both $D$ and $g_{\alpha\beta}$ are the
following:
\begin{itemize}
\item The time-like congruence $\cal R$ defined by the parametric
equations
$x^i=const.$ is, at the required approximation, a quo-harmonic congruence
and
the $x^i$
are the space quo-harmonic coordinates adapted to $\cal R$. That is to
say we assume that we have:
 
\begin{equation}
\label {1.3}
\partial_i\alpha^i_j-\frac12\partial_j\alpha=0, \quad 
\alpha=\delta^{ij}\alpha_{ij}
\end{equation}
This follows from making explicit that the functions
$x^i$ themselves are solutions of Eqs. \ref {0.15bis}.
 
\item The foliation $\cal S$ defined by the hyper-surfaces $t=const.$ is
the chorodesic synchronization of $\cal R$
with seed on a world-line $W$ with parametric
equations $x^i=x^i_0=const.$. To see what this means at the required
approximation let us follow the steps below:
 
\begin{itemize}
\item Since by definition (See Sect. 2) on $W$ $t$ is
its proper time we have:
 
\begin{equation}
\label {6.12}
g_{00} \sim 1
\end{equation}
where here and in the remaining of this paragraph $\sim$ means that the
corresponding relation is satisfied on $W$.
 
\item Since by construction the chorodesics at any event $P$ of $W$
remain on a hypersuface $x^0=const$, their tangent vector have a zero
time-component, $p^0=0$. On the other hand the unit tangent vector at
$P$ to $W$, with components $u^0=1,u^i=0$, is orthogonal to every
chorodesic through $P$ therefore we shall have:
 
\begin{equation}
\label {6.13}
g_{0i} \sim 0
\end{equation}
 
\item From $p^0=0$ all along the path of every chorodesic passing
through $P$ and from \ref {0.53} it follows also that:   
 
\begin{equation}
\label {6.14}
\frac {d }{d\lambda}p^0\sim 0\quad \hbox{or}\quad
\Gamma^{\prime 0}_{ij}p^ip^j\sim 0
\end{equation}
Using the explicit expressions of $\Gamma^0_{ij}$ and $\Sigma_{ij}$ as
functionals of $g_{\alpha\beta}$, and taking into
account that the relations above have to hold whatever the values of
$p^i$, we readily find that:
 
\begin{equation}
\label {6.15}
\partial_i g_{0j}+\partial_j g_{0i} \sim 0
\end{equation}
\end{itemize}
 
This process could be continued indefinitely to prove that whatever the
number of derivatives one has:
 
\begin{equation}
\label {6.16}
\partial_{(ijk\cdots}g_{s)0}\sim 0
\end{equation}
the round brackets meaning complete symmetrization. But to implement
this construction at the post-Newtonian level only the three first
steps are necessary.

\item $f_{0i}$ in \ref {1.1} will be required to be a linear function of
$x^i$ so that the rotation rate of $\cal R$ at the order $1/c$
be always a function of time only. This will allows to include in our
framework the general theory of Newtonian frames of reference.
 
\begin{equation}
\label {1.5}
\partial_if_{0j}-\partial_jf_{0i}=\omega_{ij}(t)
\end{equation}

\end{itemize}
 
This general framework will be called {\it The post-Newtonian
formalism}. It can be further restricted to deal with more
specific situations corresponding to two broad categories:
 
\begin{itemize}
\item The first one is that for which the Riemann tensor of the
space-time metric is zero at the appropriate approximation, this
meaning here that:
 
\begin{equation}
\label {1.6}
R_{0i0j}=O(c^{-6}), \quad R_{0ijk}=O(c^{-5}), \quad
R_{ijkl}=O(c^{-4})
\end{equation}
 
in which case \ref {1.1} is an approximation to an inertial field.
 
\item the second broad category includes all cases for which
\ref {1.6} does not hold, and among them all metric theories of
gravitation, alone or coupled to other fields. No use whatsoever of
specific field equations will be made in this paper that will cover
also therefore any appropriate parameterized formalism.
 
\end{itemize} 
 
Two particular cases have to be mentioned because the general
framework starts from them. They are the following:
 
\begin{itemize}
\item  Inertial and gravitational fields at the Newtonian
approximation can be described assuming that:

\begin{equation}
\label {1.7}
h_{\alpha\beta}=0
\end{equation}
 
\item An extended Newtonian theory of
gravitation that we discussed in \cite{Bel90} can be described
assuming that:
 
\begin{equation}
\label {1.8}
h_{00}=0, \quad h_{ij}=0
\end{equation}
\end{itemize}
Because in both cases $h_{ij}=0$ these two approximate\, \footnote{In
\cite {Bel90} both theories where described as exact. Newton's theory
as an affine theory and the extended theory as a semi-metric one.}
frameworks can be said, in a generalized sense,
to be invariant under the group of rigid motions:
 
\begin{equation}
\label {1.9}
t=t, \quad x^i=s^i(t)+R^i_j(t)y^i
\end{equation}
where $R^i_j(t)$ is a time-dependent rotation matrix. It is this
family of transformations that we generalize in the two next
sections.
 
At the post-Newtonian approximation the Newtonian field is:
 
\begin{eqnarray}
\label {1.10}
\Lambda_i&=&\lambda_i + \frac{1}{c^2}\Xi_i \\
\lambda_i&=&-\frac12\partial_i f_{00}-\partial_t f_{0i} \\
\Xi_i&=&-(\frac12\partial_i h_{00}+\frac32\partial_t f_{00}f_{0i}
+\partial_t h_{0i} -f_{00}\lambda_i)
\end{eqnarray}
 
The Coriolis, or rate of rotation field, is:
 
\begin{eqnarray}
\label {1.11}
\Omega_{ij}&=&\omega_{ij}+\frac{1}{c^2}\Psi_{ij}\\
\omega_{ij}&=&\partial_i f_{0j}-\partial_j f_{0i}\\
\Psi_{ij}&=&\partial_i h_{0j}-\partial_j h_{0i}
+\frac12 f_{00}\omega_{ij} \nonumber \\
& &-(2\lambda_i+3\partial_t f_{0i})f_{0j}
+(2\lambda_j+3\partial_t f_{0j})f_{0i}
\end{eqnarray}
 
And the so-called rate of deformation field is:
 
\begin{equation}
\label {1.12}
\Sigma_{ij}=\frac{1}{c^2}\partial_t \alpha_{ij}
\end{equation}
 
\section{Post-Newtonian quo-harmonic coordinate transformations}
 
We assume in this section that the potentials \ref {1.1}
considered in Sect. 2 are defined in the appropriate global domain
and are referred to a frame of reference of a particular type that we
shall call {\it quo-inertial}. By definition this type is defined as
equivalent to the class of frames of reference satisfying the
following conditions:
 
\begin{itemize}
\item The quo-harmonic congruence defining the meta-rigid motion of
the frame of reference contains one or many time-like geodesics.
 
\item The restricted {\it Principle of geodesic equivalence} can be
implemented on at least one of these geodesics, say $\cal C$, to be
called
the center of motion of the frame of reference. This meaning that a
system of global adapted quo-harmonic coordinates can be found such that
on
this geodesic the Fermat tensor is the unit tensor, and the Zel'
manov-Cattaneo connection is zero.

\begin{equation}
\label {2.39}
\hat g_{ij}(x^s_0)=\delta_{ij}, \quad \hat\Gamma^i_{jk}(x^s_0)=0
\end{equation}  
$x^i_0$ being the coordinates of $\cal C$. As we shall see this is
locally always possible. It is an assumption to say that a globally
well-defined quo-harmonic congruence has this property.
\end{itemize}
 
We consider below the construction of any other quo-inertial frame of
reference as a second order approximate development around its
center world-line. This construction can be split in two steps:
 
\begin{itemize}
\item The first step defines the meta-rigid motion by giving the
parameterized equations of a quo-harmonic congruence
referring this congruence to a system of quo-harmonic coordinates.
 
\item The second step constructs the chorodesic synchronization
seeded on the center geodesic of the quo-inertial frame of reference.
 
\end{itemize}
 
As already mentioned in Sect. 2, these two steps have very different
physical meanings and geometrical interpretations. The first step has
to do with a modeling of an idealized rigid body with its motion
being referred to the global frame of reference. The second step in
making a particular choice of a possible time-distribution protocol.
Both are necessary to be able to define physically meaningful tensor
transformations between the two frames of reference.
 
Let us consider a geodesic with parameterized equations:
 
\begin{equation}
\label {2.40}
x^i=s^i(t)
\end{equation}
To implement the construction of a quo-inertial meta-rigid motion
centered on this geodesic we consider the congruence defined by
equations:
 
\begin{eqnarray}
\label {2.3}
t&=&t \\
x^i&=&s^i(t)+R^i_p(t)y^p+\frac{1}{c^2}R^i_p(t)\delta_2y^p \\
\delta_2y^p(t)&=&L^p_r(t)y^r+\frac{1}{2}Q^p_{rk}(t)y^ry^k
+\frac{1}{3!}C^p_{rkl}(t)y^ry^ky^l
\end{eqnarray}
where $R^i_p(t)$ is a time dependent rotation matrix and $y^i$ can
equivalently be considered either as parameters or as a system of
adapted coordinates of the new congruence. Obviously the zero order
choice in these expressions has been chosen to guarantee the correct
Newtonian limit. Also to simplify the writing we have chosen the
coordinates $y^i$ such that the parametric equations of the geodesic
$\cal C$ of the new congruence be $y^i=0$. 
 
Differentiating the expressions above we obtain:
 
\begin{equation}
\label {2.4}
dt=dt, \quad dx^i=W^idt+R^i_jdy^j +\frac{1}{c^2}(A^i_kdy^k+B^idt),
\end{equation}
where:
 
\begin{eqnarray}
\label {2.6}
W^i&=&v^i+\dot R^i_py^p \\
A^i_j&=&R^i_p(L^p_j+Q^p_{jk}y^k+\frac{1}{2}C^p_{jkl}y^ky^l) \\
B^i&=&R^i_p(\dot L^p_ry^r+\frac{1}{2}\dot Q^p_{rk}y^ry^k
+\frac{1}{3!}\dot C^p_{rkl}y^ry^ky^l)
\end{eqnarray}
where $v^i=\dot s^i$, the over-head dot meaning a derivative with respect
to $t$.
 
A straightforward calculation gives the transformed basic metric
quantities: 
 
\begin{eqnarray}
\label {2.5}
\bar g_{00}&=&-1+\frac{1}{c^2}\bar f_{00}
+\frac{1}{c^4}\bar h_{00} \\
\bar f_{00}&=&f_{00}+2f_{0i}W^i+W^2 \\
\bar h_{00}&=&h_{00}+2h_{0i}W^i+2(f_{0i}+W_i)B^i+h_{ij}W^iW^j+h^*_{00} \\
h^*_{00}&=&+(\partial_kf_{00}
+\omega_{ki}R^k_pW^i)\delta_2y^p
\end{eqnarray}

\begin{eqnarray}
\label {2.7}
\bar g_{0i}&=&\frac{1}{c}\bar f_{0i}
+\frac{1}{c^3}\bar h_{0i} \\
\bar f_{0i}&=&(f_{0s}+W_s)R^s_i \\
\bar h_{0i}&=&(h_{0s}+B_s+h_{sk}W^k)R^s_i +(f_{0k}+W_k)A^k_i
+h^*_{0i} \\
h^*_{0i}&=&\frac12\omega_{rs}R^r_kR^s_i\delta_2y^k
\end{eqnarray}

\begin{eqnarray}
\label {2.8}
\bar g_{ij}&=&\delta_{ij}+\frac{1}{c^2}\bar h_{ij} \\
\bar h_{ij}&=&h_{rs}R^r_iR^s_j+\delta_{rs}(A^r_iR^s_j+A^r_jR^s_i), \quad
\end{eqnarray}
 
\begin{eqnarray}
\label {2.8bis}
\hat{\bar g}_{ij}&=&\delta_{ij}+\frac{1}{c^2}\bar\alpha_{ij} \\
\bar\alpha_{ij}&=&\bar h_{ij}+\bar f_{0i}\bar f_{0j}
\end{eqnarray}
the starred terms in \ref {2.5} and \ref {2.7} coming from
substituting $x^i$ in $\bar f_{00}$ and $\bar f_{0i}$ according to
\ref {2.3}.
 
The congruence defined by the parametric Eqs. \ref {2.3} will be a
quo-harmonic one if we choose the free functions of time $L^i_j(t),
Q^p_{rk}(t)$ and $C^p_{rkl}(t)$ in such way that:
 
\begin{equation}
\label {2.43}
\frac{\partial }{\partial y^i}\bar\alpha^i_j
-\frac12\frac{\partial }{\partial y^j}\bar\alpha =0
\end{equation}
i.e., in such a way that the parameters $y^i$ are quo-harmonic
coordinates.
 
The conditions stated in \ref {2.39} are further restrictions on these
conditions. We shall rewrite them at the appropriate approximation
as:
 
\begin{equation}
\label {2.17}
\alpha_{ij}\sim 0, \quad
\frac{\partial }{\partial y^k}\bar\alpha_{ij} \sim 0
\end{equation}
where from now on the symbol $\sim$ will mean that the coordinates
$x^i$ have been replaced by $s^i(t)$ and $y^k$ by $0$.
 
Taking into account \ref {2.7} - \ref {2.8bis} the first group of
conditions require that:
 
\begin{equation}
\label {2.42}
L_{(ij)} \sim -\frac12(h_{rs}+(f_{0s}+v_s)(f_{0r}+v_r))R^s_iR^r_j
\end{equation}
where:
 
\begin{equation}
\label {2.27}
L_{ij}=\delta_{ik}L^k_j, \quad  L_{(ij)}=\frac12(L_{ij}+L_{ji})
\end{equation} 
We shall see at the end of this section that the antisymmetric part of
$L_{ij}$ can be chosen as to give an unambiguous physical meaning to the
rotation matrix $R^i_j$.
 
The second group of conditions
\ref {2.39} leads to:
 
\begin{equation}
\label {2.28}
Q_{j,ik}+Q_{i,jk}+X_{k,ij} \sim 0
\end{equation}
or:
 
\begin{equation}
\label {2.30}
Q_{i,jk}=\frac12(X_{j,ki}+X_{k,ji}-X_{i,jk})
\end{equation}
where:
 
\begin{eqnarray}
\label {2.29}
X_{k,ij}&=&[\partial_lh_{rs}+\frac12(\omega_{ls}(f_{0r}+v_r)
+(f_{0s}+v_s)\omega_{lr}]R^l_kR^r_iR^s_j \\
& &+\frac12(\Delta_{ki}(f_{0r}+v_r)R^r_j+\Delta_{kj}(f_{0s}+v_s)R^r_i)
\end{eqnarray}
and:
 
\begin{equation}
\label {2.29bis}
\Delta_{ij}=2\dot R^s_iR_{sj}
\end{equation}
Notice that the conditions \ref {2.17} could have been implemented on
any world-line. But it is only because we have assumed that the
world-line $x^i=s^i(t)$ is a geodesic that these conditions are
physically justified and required by the principle of geodesic
equivalence.
 
Consider now the next step in the approximation. Using again
\ref {2.7} - \ref {2.8bis} we have:
 
\begin{equation}
\label {2.31}
\frac{\partial^2   }{\partial y^ky^l}\bar\alpha_{ij} \sim
Y_{ij,kl}+C_{i,jkl}+C_{j,ikl}
\end{equation}
where:
 
\begin{eqnarray}
\label {2.32}
Y_{ij,kl}&=&\partial^2_{mn}\alpha_{rs}R^r_iR^s_jR^m_kR^n_l \\
&
&+\frac14(\bar\omega_{ki}\bar\omega_{lj}+\bar\omega_{kj}\bar\omega_{li})
-\frac14(\tilde\omega_{ki}\tilde\omega_{lj}
+\tilde\omega_{kj}\tilde\omega_{li})
\end{eqnarray}
where:
 
\begin{equation}
\label {2.35}
\bar\omega_{ij}=\tilde\omega_{ij}+\Delta_{ij}, \quad
\tilde\omega_{ij}=\omega_{rs}R^r_iR^s_j
\end{equation}
At the order $1/c^2$ we have:
 
\begin{equation}
\label {2.44}
\hat{\bar R}_{ijkl} \sim \hat{\tilde R}_{ijkl}
+\frac34(\bar\omega_{ij}\bar\omega_{kl}
-\tilde\omega_{ij}\tilde\omega_{kl})
\end{equation}
where:
 
\begin{eqnarray}
\label {2.45}
\hat{\tilde R}_{ijkl}&=&-\frac12
(\partial^2_{mn}\alpha_{rs}+\partial^2_{rs}\alpha_{mn}
-\partial^2_{mr}\alpha_{ns}-\partial^2_{ns}\alpha_{mr})
R^r_iR^s_jR^m_kR^n_l \\
\hat{\bar R}_{ijkl}&=&-\frac12
(\frac{\partial^2   }{\partial y^iy^k}\bar\alpha_{jl}+
\frac{\partial^2   }{\partial y^jy^l}\bar\alpha_{ik}
-\frac{\partial^2   }{\partial y^iy^l}\bar\alpha_{jk}
-\frac{\partial^2   }{\partial y^jy^k}\bar\alpha_{il})
\end{eqnarray}
Since \ref {2.44} is independent of $C^i_{jkl}$ and $\Delta_{ij}$ is
an arbitrary skew-symmetric quo-tensor, it is not in general possible
to nullify the second derivatives of the Fermat tensor on the center
of a quo-inertial motion.
 
On the contrary it is always possible to implement the second order
quo-harmonic condition:
 
\begin{equation}
\label {2.33}
\frac{\partial }{\partial y^k}
(\frac{\partial }{\partial y^i}\bar\alpha^i_j
-\frac12\frac{\partial }{\partial y^j}\bar\alpha) \sim 0
\end{equation}
In fact from \ref {2.31} and \ref {2.32} it follows that the preceding
condition is equivalent to:

\begin{equation}
\label {2.46}
C_{ijkl}\delta^{kl} \sim \frac12(\bar\omega_{ik}\bar\omega_{jl}
-\tilde\omega_{ik}\tilde\omega_{jl})\delta^{kl}
\end{equation}
The solution of this equation compatible with the symmetry of the
problem is:
 
\begin{equation}
\label {2.34}
C^s_{ijk}=\frac14(\bar\Theta^s_i\bar\delta_{jk}
+\bar\Theta^s_j\bar\delta_{ki}
+\bar\Theta^s_k\bar\delta_{ij})
-\frac14(\tilde\Theta^s_i\tilde\delta_{jk}
+\tilde\Theta^s_j\tilde\delta_{ki}
+\tilde\Theta^s_k\tilde\delta_{ij})
\end{equation}
where:
 
\begin{equation}
\label {2.36}
\bar\Theta_{ij}=\frac12\epsilon^k\bar\omega_{ik}\bar\omega_{kj}, \quad
\tilde\Theta_{ij}=\frac12\epsilon^k\tilde\omega_{ik}\tilde\omega_{kj},
\quad
\epsilon^k=1
\end{equation}

\begin{equation}
\label {2.37}
\bar\delta_{jk}=-2\bar\Theta_{jk}/(\bar\omega_{rs}\bar\omega^{rs})  
\end{equation}

\begin{equation}
\label {2.38}
\tilde\delta_{jk}=-2\tilde\Theta_{jk}/(\tilde\omega_{rs}\tilde\omega^{rs})
\end{equation}
 
To understand the meaning of this solution let us look at the
coordinate
transformations \ref {2.3} as the composition of two transformations:
 
\begin{eqnarray}
\label {2.50}
x^i&=&s^p+R^{\prime i}_p(y^{\prime p}+\frac{1}{c^2}
(L^{\prime p}_r(t)y^{\prime r}
+\frac{1}{2}Q^{\prime p}_{rk}(t)y^{\prime r}y^{\prime k})
+\frac{1}{3!}C^{\prime p}_{jkl}y^{\prime j}y^{\prime k}y^{\prime
l})) \\
y^{\prime i}&=&R^{\prime\prime i}_p(y^p
+\frac{1}{3!c^2}C^{\prime\prime p}_{jkl}y^jy^ky^l)
\end{eqnarray}
with:
 
\begin{equation}
\label {2.52}
R^{\prime i}_kR^{\prime\prime k}_j=R^i_j
\end{equation}
which leads to the composition law:
 
\begin{equation}
\label {2.55}
C^i_{jkl}=C^{\prime\prime i}_{jkl}
+\tilde R^{\prime i}_pC^{\prime p}_{mnr}R^{\prime m}_jR^{\prime n}_k
R^{\prime r}_l
\end{equation}
If we choose $R^{\prime i}_k$ such that:
 
\begin{equation}
\label {2.54}
2\dot R^{\prime k}_i=-\omega_{mk}R^{\prime m}_i
\end{equation}
i.e. if choose $R^{\prime i}_k$ such that, according to
\ref {2.29bis} and \ref {2.35}, the rotation rate of the congruence
defined by the first group of equations \ref {2.50} is zero, then from
\ref {2.34} we have:
 
\begin{equation}
\label {2.56}
C^{\prime i}_{jkl}=-\frac14(\Theta^{\prime s}_i\delta^\prime_{jk}
+\Theta^{\prime s}_j\delta^\prime_{ki}
+\Theta^{\prime s}_k\delta^\prime_{ij})
\end{equation}
with:
 
\begin{equation}
\label {2.57}
\Theta^\prime_{rs}=
\frac12\epsilon^k\omega^\prime_{ik}\omega^\prime_{kj}, \quad
\delta^\prime_{jk}=
-2\Theta^\prime_{jk}/(\omega^\prime_{rs}\omega^{\prime rs}), \quad
\omega^\prime_{rs}=\omega_{ij}R^{\prime i}_rR^{\prime j}_s
\end{equation}
this being the unique acceptable solution of \ref {2.46} if
$\bar\omega^\prime_{ij}=0$, because one has to take into account the
complete symmetry of $C^{\prime i}_{jkl}$ with respect to its
covariant indices and the facts that $\Theta^\prime_{ij}$ is symmetric
and that the dual of $\omega^\prime_{ij}$ is an eigen-vector with
eigen-value zero.
 
Using again \ref {2.46} and \ref {2.34} with
$\tilde\omega^\prime_{ij}=0$, we get:
 
\begin{equation}
\label {2.58}
C^{\prime\prime s}_{ijk}=\frac14(\bar\Theta^{\prime
s}_i\bar\delta^\prime_{jk}
+\bar\Theta^{\prime s}_j\bar\delta^\prime_{ki}
+\bar\Theta^{\prime s}_k\bar\delta^\prime_{ij})
\end{equation}
 
Combining \ref {2.52}, \ref {2.55}, \ref {2.57} and \ref {2.58} we
uncover the
meaning of
$R^{\prime i}_j$ and $R^{\prime\prime i}_j$. The first is the
rotation leading to the non rotating congruence centered on the
geodesic of the transformed congruence; the second  is the
rotation of the latter congruence as measured from the non rotating
reference defined by the first one.
 
To determine the skew-symmetric part of $L_{ij}$ we shall demand on the
center geodesic $C$ of the new congruence that:
 
\begin{equation}
\label {2.61}
\bar\Psi_{ij} \sim 0
\end{equation}
i.e. that the post-Newtonian correction to $\omega_{ij}$ on $C$ be zero.
This
requirement does not have to be considered as an additional demand to
the quo-harmonic condition but as a precision made to the meaning of
the matrix $R^i_j$ in the transformations \ref {2.3}. Equivalently we
could say that the real parameter in these transformations is not
$R^i_j$ but the rotation rate $\bar\omega_{ij}$, on $\cal C$, of the
new congruence. In fact considering the latter as a free parameter it
does not make sense to consider separately its classical and
relativistic correction. It has to be considered as a whole and have
the meaning that it has at the classical approximation. 
 
From \ref {1.11} and from \ref {2.6} - \ref{2.8} it follows that
$\bar\Psi_{ij}$ 
can be written as:
 
\begin{equation}
\label {2.60}
\bar\Psi_{ij} \sim \dot
L_{[ij]}+\bar\omega_{ik}L^k_j-\bar\omega_{jk}L^k_i
+ Z_{ik}, \quad L_{[ij]}=\frac12(L_{ij}-L_{ji})
\end{equation}
where both the symmetric part of $L_{ij}$, as given by \ref {2.42},
and the skew-symmetric
object $Z_{ij}$ depend only on quantities that have been already
calculated. It follows that the requirement \ref {2.61}
is equivalent to a differential equation for $L_{[ij]}$ whose general
solution will depend on its value at a given instant. These constants
can be considered to be zero without any loss of generality because
whatever their value they will all lead to the same congruence.
 
\section{Post-Newtonian chorodesic synchronizations}
 
We consider in this section a chorodesic foliation $\cal F$ orthogonal
to the center geodesic $C$ of the meta-rigid motion we are
considering, i.e. the family of hyper-surfaces spanned by the
chorodesics orthogonal to this geodesic.  The corresponding {\it IAT}
will be a time coordinate $t^\prime=t^\prime(t,y^i)$ such that
$t^\prime=const$ be the local equations of the chorodesic leaves of the
foliation.
 
Let $t^\prime$ be defined as the inverse of the following time
transformation:
 
\begin{eqnarray}
\label {6.1}
t&=&t^\prime+\frac{1}{c^2}\delta_2t^\prime+\frac{1}{c^4}\delta_4t^\prime
\\
\delta_2t^\prime&=&I(t^\prime)+L_i(t^\prime)y^i
+\frac12 Q_{ij}(t^\prime)y^iy^j \\
\delta_4t^\prime&=&H(t^\prime)+K_i(t^\prime)y^i
+\frac12 P_{ij}(t^\prime)y^iy^j
\end{eqnarray}
Differentiating this expression we get:
 
\begin{equation}
\label {6.2}
dt=dt^\prime+\frac{1}{c^2}(Ddt^\prime+C_idy^i)
+\frac{1}{c^4}(Fdt^\prime+D_idy^i)
\end{equation}
where:
 
\begin{eqnarray}
\label {6.3}
D&=&\dot I^\prime+\dot L^\prime_iy^i+\frac12 \dot Q^\prime_{ij}y^iy^j \\
C_i&=&L_i+Q_{ij}y^j \\
F&=&\dot H^\prime+\dot K^\prime_iy^i
+\frac12 \dot P^\prime_{ij}y^iy^j \\
E_i&=&K_i+P_{ij}y^j
\end{eqnarray}
where a dotted and primed quantity means that it has been
differentiated with respect to $t^\prime$. With these notations the
new potentials are:
 
\begin{eqnarray}
\label {6.4}
g^\prime_{00}&=&-1+\frac{1}{c^2}f^\prime_{00}
+\frac{1}{c^4}h^\prime_{00} \\
f^\prime_{00}&=&\bar f_{00}-2D \\
h^\prime_{00}&=&\bar h_{00}+2\bar f_{00}D-2F-D^2+h^{\prime *}_{00} \\
h^{\prime *}_{00}&=&\partial_t \bar f_{00}\delta_2t^\prime
\end{eqnarray} 
 
\begin{eqnarray}
\label {6.5}
g^\prime_{0i}&=&\frac{1}{c}f^\prime_{0i}
+\frac{1}{c^3}h^\prime_{0i} \\
f^\prime_{0i}&=&\bar f_{0i}-C_i \\
h^\prime_{0i}&=&\bar h_{0i}+\bar f_{00}C_i
+\bar f_{0i}D-E_i +h^{\prime *}_{0i} \\
h^{\prime *}_{0i}&=&\partial_t \bar f_{0i}\delta_2t^\prime
\end{eqnarray}
the starred terms in $h^\prime_{00}$ and $h^\prime_{0i}$ coming from
using \ref {6.1} in the lower order terms $f^\prime_{00}$ and
$f^\prime_{0i}$.
And also:
 
\begin{eqnarray}
\label {6.6}
g^\prime_{ij}&=&\delta_{ij}+\frac{1}{c^2}h^\prime_{ij} \\
h^\prime_{ij}&=&\bar h_{ij}+\bar f_{0i}C_j+\bar f_{0j}C_i-C_iC_j
\end{eqnarray}
and:
 
\begin{equation}
\label {6.7}
\hat g^\prime_{ij}=\delta_{ij}+\frac{1}{c^2}\alpha^\prime_{ij},
\quad \alpha^\prime_{ij}=\bar\alpha_{ij}
\end{equation}
this last result following directly from the fact that the Fermat
object, $\hat g_{ij}$, is a quo-tensor, i.e. a three dimensional
tensor under coordinate transformations that leave invariant the
congruence.
 
The conditions to be demanded to \ref {6.1} were derived in Sect. 3.
Taking into account \ref {6.4}, \ref {6.12} requires that:
 
\begin{eqnarray}
\label {6.8}
\dot I^\prime &\sim& \frac12 \bar f_{00} \\
\dot H^\prime &\sim& \frac12 \bar h_{00}+\frac34\bar f^2_{00}
\end{eqnarray} 
The functions
$I$ and $H$ have to be obtained integrating these differential
equations and each will depend on an arbitrary constant.
The second group of conditions \ref {6.13} and \ref {6.5} require that:
 
\begin{eqnarray}
\label {6.9}
L_i&\sim & \bar f_{0i} \\
K_i&\sim & \bar h_{0i}+\frac12 \bar f_{00}(f_{0i}+L_i)
\end{eqnarray}
Finally the third group of conditions \ref {6.15} and the relations
obtained differentiating \ref {6.5} with respect to $y^i$ yield:
 
\begin{eqnarray}
\label {6.10}
Q_{ij} & \sim &\frac12(\bar\partial_i\bar f_{0j}
+\bar\partial_j\bar f_{0i}) \\
P_{ij} & \sim &\frac12(N_{ij}+N_{ji})
\end{eqnarray}
with:
 
\begin{equation}
\label {6.11}
N_{ij}=\bar\partial_i\bar h_{0j}+\bar\partial_i\bar f_{00}L_i
+\frac12\bar f_{00}Q_{ij}+\dot L^\prime_i(\bar f_{0j}-L_j)
\end{equation}
This concludes our modelisation of the chorodesic atomic time
distribution protocol at the post-Newtonian approximation.

\section*{Appendix: Some definitions and notations}
Given any space-time with line element:
 
\begin{displaymath}
\label {7.1}
ds^2=g_{\alpha\beta}(x^\rho)dx^\alpha dx^\beta, \quad
\alpha,\beta,\cdots=0,1,2,3, \quad x^0=ct,
\end{displaymath}
let us consider a time-like congruence $\cal R$, $u^\alpha$ being its
unit
tangent vector field ($u_\alpha u^\alpha=-1$). The {\it Projector} into
the plane orthogonal to $u^\alpha$ is:
 
\begin{displaymath}
\label{7.14}
\hat g_{\alpha \beta }=g_{\alpha \beta }+u_\alpha u_\beta 
\end{displaymath}
 
By definition the {\it Newtonian field} is the opposite to the
acceleration field:
 
\begin{displaymath}
\label {7.10}
\Lambda_\alpha=-u^\rho \nabla _\rho u_\alpha,
\quad \Lambda_\alpha u^\alpha=0.
\end{displaymath}
The {\it Coriolis field}, or the rotation rate field, is the
skew-symmetric
2-rank tensor orthogonal to $u^\alpha $:
 
\begin{displaymath}
\label{7.11}
\Omega _{\alpha \beta }=\hat \nabla _\alpha u_\beta -\hat
\nabla _\beta u_\alpha ,\quad \Omega _{\alpha \beta }u^\alpha =0.
\end{displaymath}
where:
 
\begin{displaymath}
\label{7.11bis}
\hat \nabla _\alpha u_\beta \equiv \hat g_\alpha ^\rho \hat g_\beta
^\sigma
\nabla _\alpha u_\beta .
\end{displaymath}
And {\it Born's deformation rate field} is the symmetric 2-rank
tensor orthogonal to $u^\alpha $:
 
\begin{displaymath}
\label{7.12}
\label{deformation}\Sigma _{\alpha \beta }=\hat \nabla _\alpha u_\beta
+\hat
\nabla _\beta u_\alpha ,\quad \Sigma _{\alpha \beta }u^\alpha =0,
\end{displaymath}
 
Let $x^\alpha$ be a system of adapted coordinates,
i.e., such that:
 
\begin{displaymath}
\label {7.2.0}
u^i=0, \quad i,j,\cdots =1,2,3
\end{displaymath}
We use the following notations:
 
\begin{displaymath}
\label {7.2}
\xi = \sqrt{-g_{00}}, \qquad  \varphi_i = \xi^{-2}g_{0i} ,
\end{displaymath}
and:
 
\begin{displaymath}
\label {7.3}
\hat g_{ij} = g_{ij} + \xi^2\varphi_i\varphi_j, \quad
\hat g^{ij} = g^{ij}
\end{displaymath}
which we shall call the {\it Fermat} quo-tensor of the congruence.
Here and below quo-tensor refers to an object, well defined on the
quotient manifold ${\cal V}_3={\cal V}_4/\cal R$,
whose covariant components are
the space components of a tensor of ${\cal V}_4$ orthogonal to
$u^\alpha$.
 
The {\it Newtonian field} is then the quo-vector:
 
\begin{displaymath}
\label {7.4}
\Lambda_i = -c^2(\hat\partial_i \ln\xi +\frac{1}{c}\partial_t \varphi_i),
\quad
\hat\partial_i\cdot \equiv
\partial_i\cdot + \frac{1}{c}\varphi_i\partial_t\cdot
\end{displaymath}
The {\it Coriolis field}, or Rotation rate field, is the skew-symmetric
quo-tensor:
 
\begin{displaymath}
\label {7.7}
\Omega_{ij}=c\xi(\hat\partial_i \varphi_j-\hat\partial_j \varphi_i)
\end{displaymath}
and the {\it Born's deformation rate field} is the symmetric quo-tensor:
 
\begin{displaymath}
\label {7.5}
\Sigma_{ij}=\hat\partial_t\hat g_{ij},
\quad \hat\partial_t=\xi^{-1}\partial_t
\end{displaymath}
 
To these familiar geometrical objects it is necessary
to add the following ones, \cite{Zelmanov},
\cite{Cattaneo}, \cite{Ida}:
 
\begin{displaymath}
\label {7.8}
\hat\Gamma^i_{jk} = \frac12 \hat g^{is}(\tilde\partial_j\hat g_{ks}
+ \tilde\partial_k\hat g_{js} -\tilde\partial_s\hat g_{jk}) ,
\end{displaymath}
which are the {\it Zel'manov-Cattaneo symbols}, and:
 
\begin{displaymath}
\label {7.9}
\hat R^i_{jkl} = \tilde\partial_k\hat\Gamma^i_{jl} -
\tilde\partial_l\hat\Gamma^i_{jk} + \hat\Gamma^i_{sk}\hat\Gamma^s_{jl} -
\hat\Gamma^i_{sl} \hat\Gamma^s_{jk}
\end{displaymath} 
which is the {\it Zel'manov-Cattaneo quo-tensor},

\end{document}